\documentclass[a4paper,12pt]{article}

\usepackage{graphicx,epsf,color}
\usepackage{amssymb}

\newcommand{\rme}{\mathrm{e}}

\newcommand{\rmd}{\mathrm{d}}

\newcommand{\ad }{a^{\dagger}}

\newcommand{\op}{ \oplus  }
\newcommand{\om}{ \ominus}

\newcommand{\ep}{\epsilon}

\newcommand{\lb }{ \left (}
\newcommand{\rb }{ \right )}

\begin{document}

\begin{center}
{\Large{\textbf{Photon Gas at the Planck Scale within\\ \vskip10pt the Doubly Special Relativity}}}

\vspace*{0.5cm}
{W.S. Chung${}^{a,b}$, A.M. Gavrilik${}^{c,d}$, A.V. Nazarenko${}^{c,e}$}

\vspace*{0.5cm}
{\it${}^a$Department of Physics and Research Institute of Natural Science,\\
 College of Natural Science,\\
 Gyeongsang National University, Jinju 660-701, Korea\\
${}^b$mimip44@naver.com\\
${}^c$Bogolyubov Institute for Theoretical Physics of NAS of Ukraine,\\
14-b, Metrolohichna str., Kyiv 03143, Ukraine\\
${}^d$omgavr@bitp.kiev.ua\\
${}^e$nazarenko@bitp.kiev.ua}

\end{center}

\begin{abstract}
Within the approach to doubly special relativity (DSR) suggested by
Magueijo and Smolin, a new algebraically justified rule of so-called
$\kappa$-addition for the energies of identical particles is
proposed. This rule permits to introduce the nonlinear
$\kappa$-dependent Hamiltonian for one-mode multi-photon
(sub)system. On its base, with different modes treated as
independent, the thermodynamics of black-body radiation is explored
within DSR, and main thermodynamic quantities are obtained. In their
derivation, we use both the analytical tools within mean field
approximation (MFA) and numerical evaluations based on exact
formulas. The entropy of one-mode subsystem turns out to be finite
(bounded). Another unusual result is the existence of threshold
temperature above which radiation is present. Specific features of
the obtained results are explained and illustrated with a number of
plots. Comparison with some works of relevance is given.


\end{abstract}

\newpage
\section{Introduction}

Recently, the deformation of the  special relativity was
accomplished so that it might jointly admit two invariant
fundamental scales: one of them being the speed of light, and the
other -- an energy scale naturally identifiable with Planck energy
\cite{AC02,AC01,LNRT,LNR,MS02}. This deformed theory is usually
called a doubly special relativity (DSR) because it has two
invariant scales. The DSR is important when one aims to describe the
particles' dynamics at very high energies approaching the Planck
scale. It is believed that at this scale the spacetime structure can
be influenced by effects of quantum gravity.

Initially DSR was introduced in an algebraic way, using certain
quantum deformations of the Lorentz group \cite{LNRT,LNR}. Later it
was also derived by taking as starting point a few physical
postulates, including the requirement that it should reduce
to special relativity if the low-energy limit is applied~\cite{AC02,AC01}.

In the ordinary special relativity, the momentum changes under the
Lorentz transformation with the infinitesimal
boost generator $J$ as
\begin{equation}
\delta p_0 = \{ J, p_0 \} =p_1, \qquad
\delta p_1 = \{J, p_1\} = p_0,
\end{equation}
where we consider  1+1 dimensions,  and $\{, \}$ denotes the
Poisson bracket. The DSR model by Magueijo--Smolin (MS model)
\cite{MS02} takes the form
\begin{equation}\label{trans2}
\delta p_0 = \{ J, p_0 \} = \lb 1 - \frac{p_0}{\kappa}\rb p_1, \qquad
\delta p_1 = \{J, p_1\} = p_0 -\frac{p_1^2}{\kappa},
\end{equation}
with $\kappa$ the Planck energy. Due to the factor $1-p_0/\kappa$,
for $p_0 =\kappa$ we have $\delta p_0 =0$, and thus the invariance
of Planck energy. Under the transformation (\ref{trans2}), one has
the invariant quantity $p^2/(1-p_0/\kappa)^2,$ where $p^2=p_0^2-p_1^2$.

The DSR  models imply that the momentum of a particle transforms
nonlinearly under the Lorentz group. In this respect, there is a
variety of ways to realize nonlinear representations of the Lorentz
group obeying the above-mentioned postulates, but, to develop
the theory there are two basic routes: working in ordinary spacetime,
or to explore a spacetime involving non-commuting coordinates.
In the noncommutative case, for defining the Hamiltonian
classical dynamics, noncanonical Poisson brackets are used.
It is clear that, depending on the choice of initial assumptions,
the explored models may lead to differing physical consequences or
predictions.

In this paper we propose a new, DSR-inspired rule of
"$\kappa$-addition" for the energies of particles (Section 2), study
its properties, and then introduce the corresponding Hamiltonian for
one-mode multi-photon system.
On its base, assuming independence of different modes, we explore
in Sec.~3 the thermodynamics of black-body radiation in the
framework of DSR and derive basic thermodynamic quantities.
Our calculations lead to rather unusual results, whose special features
are analyzed and illustrated with plots. In the final section, main
consequences and conclusions are presented, along with brief comparison
with respective aspects in some related works.

\section{New Addition Rule for Energies and the Hamiltonian}

The exact dispersion relation for DSR (Magueijo-Smolin model) \cite{JV} is
\begin{equation}
\frac{E^2-p^2}{(1-E/\kappa)^2} = m^2,
\end{equation}
where $p_0 = E$ and $p = |\vec{p}|$ and $m$ is the rest mass.  This
relation is transformed into the ordinary dispersion relation
through the (direct/inverse) map
\begin{equation}
\ep = \frac{E}{1-E/\kappa}, \qquad
\vec{\pi}_i = \frac{\vec{p}_i}{1-E/\kappa},
\end{equation}
\begin{equation}
E = \frac{ \ep}{1+\ep/\kappa}, \qquad
\vec{p}_i = \frac{\vec{\pi}_i}{1+\ep/\kappa},
\end{equation}
so that we have $\ep^2-\pi^2 = m^2.$

Eq.~(5) shows that $E=\kappa$ in the limit $\ep \rightarrow \infty$,
and from $\ep \ge 0$ in Eq.~(5) we have $ E\le\kappa$, which implies
that there exists a {\it maximum energy in Nature}.

Instead of the energy conservation corresponding to the undeformed
energy, $\ep_\mathrm{tot} = \ep_1 + \ep_2,$ we propose the new
conservation of energy in DSR theory as
\begin{equation}\label{Etot}
E_\mathrm{tot} = E_1 \op_{\kappa} E_2,
\end{equation}
with the "$\kappa$-addition"
\begin{equation} \hspace{12mm}
a \op_{\kappa} b = \frac{\frac{a}{1- a/\kappa} +
\frac{b}{1-b/\kappa}}{1+ \frac{1}{\kappa}\lb
\frac{a}{1-a/\kappa}+\frac{b}{1-b/\kappa}\rb }
\end{equation}
which can also be written as
\[
\frac{1}{\frac{1}{a \op_{\kappa} b}-\frac{1}{\kappa}} =
\frac{1}{\frac{1}{a}-\frac{1}{\kappa}} +
\frac{1}{\frac{1}{b}-\frac{1}{\kappa}} \, .
\]
It is clear that $\kappa$-addition reduces to ordinary addition in
the limit $\kappa \rightarrow \infty$.


\textbf{Properties of $\kappa$-Addition:} 1) The $\kappa$-addition
is commutative: $a\op_{\kappa}b=b\op_{\kappa}a$; \ 2) it is
associative: $(a\op_{\kappa}b)\op_{\kappa} c =
 a \op_{\kappa} ( b \op_{\kappa} c)$; \ 3) there exists the $0$-identity:
 $a \op_{\kappa} 0 = a$; \
4) there exists the inverse element for $a$ denoted by $\om_{\kappa}
a$: $a \op_{\kappa} ( \om_{\kappa} a)=0,$ which gives $\om_{\kappa}
a = - \frac{a}{1-2 a/\kappa}$; \ 5) the $\kappa$-addition of
$\kappa$ and any element $a$ gives $\kappa$: $a \op_{\kappa} \kappa
= \lim_{b\rightarrow \kappa} a \op_{\kappa}b=\kappa.$ The last
property implies that Planck energy is the maximum energy: $E
\op_{\kappa} \kappa = \kappa.$ Besides, for any positive energies
$E_1$, $E_2$ \ ($\neq \kappa$) we know that $E_1 \op_{\kappa} E_2 <
\kappa$. Thus, Planck energy is the maximum 
energy as the light speed is the maximum velocity.


\textbf{Photons at the Planck Scale.} Photon obeys the deformed
dispersion relation at the Planck scale, which gives the same one as
in the undeformed theory, $E=p$.

But, the energy of photon obeys the deformed law of the energy
conservation given by $\kappa$-addition. Thus, the total energy for
$n$ photons differs from the $n$ times the energy of single photon,
$E_{n~photons} \ne n E_{1 ~photon}.$

For example, the energy for two identical photons at the Planck
scale obeys
\begin{equation}
E_{2~photons} =  E_{1~photon} \op_{k} E_{1~photon} = \frac{2 E_{1~photon}}{
 1 + E_{1~photon}/\kappa} \, .
\end{equation}
If we set
\begin{equation}
 n \odot_{\kappa} a =
 \underbrace{a \op_{\kappa} a \op_{\kappa} \cdots \op_{\kappa} a}_n
\end{equation}
the energy of $n$ photons is given as
\begin{equation}\label{En1}
E_{n~photons} = n \odot_{\kappa} E_{1~photon} = \frac{ n
E_{1~photon}}{1 + ( n-1) E_{1~photon}/\kappa}
\end{equation}
where we require $E_{1~photon}=h\nu \le \kappa$. Quite interestingly, we find
\begin{equation}
\lim_{n \rightarrow \infty} E_{n~photons}  = \kappa.
\end{equation}

Thus, the Hamiltonian for photon at Planck scale is
\begin{equation}\label{Hh}
H = \frac { N h \nu}{1+(N-1) h \nu /\kappa}
\end{equation}
where the photon creation/destruction operators obey
\begin{equation}\label{alg}
[a, \ad] =1, \qquad N=\ad a
\end{equation}
and the photon number state is taken as $N|n\rangle = n|n \rangle$,
$n=0, 1, 2, \ldots$.

 The single-mode (or monochromatic) multi-photon Hamiltonian (\ref{Hh}) coupled
 with usual bosonic formulas (\ref{alg}) principally differs from the familiar
 free or linear one $H=N h \nu$: \ being {\it essentially nonlinear}
 (with anharmonicities of all orders) it implies highly nontrivial
 self-interaction within each one mode. In other words, the imposed special
 rule of $\kappa$-addition given in (\ref{En1}), naturally induces
 nontrivial self-interaction seen in the Hamiltonian (\ref{Hh}).
 Clearly, the strength of nonlinearity is controlled by $\kappa$.


\textbf{Similarity with a Class of Deformed Oscillators.}
It is worth noting that the nonlinear Hamiltonian (\ref{Hh})
can also be obtained from a different standpoint beyond the DSR,
namely within the theory of deformed oscillators (deformed bosons).
That is, the particular deformed oscillator (DO) whose creation/destruction
operators $\tilde{a}$, $\tilde{a}^\dagger$ obey the relations
\begin{equation}\label{DSF}
 [\tilde{a},  \tilde{a}^\dagger] = \phi_{\kappa}(N+1) - \phi_{\kappa}(N), \quad
 \phi_{\kappa}(N)\equiv \tilde{a}^\dagger  \tilde{a} =
 \frac { N }{ 1 + {\kappa}^{-1}( N-1) h \nu},
\end{equation}
defined by the respective deformation structure function (DSF)
$\phi_{\kappa}(N)$ (for the concept of DSF see e.g.~\cite{Bon}),
here with the deformation parameter $h\nu/{\kappa}$. It leads to
the free Hamiltonian of deformed bosons of the form
$$
 \tilde{H}= h\nu\, \tilde{a}^\dagger \tilde{a} = h\nu\, \phi_{\kappa}(N),
$$
which coincides with that in (\ref{Hh}).
In this case, self-interaction is embodied in deformation.
Clearly, if ${\kappa}^{-1}\to 0$ both $H$ and $\tilde{H}$
reduce to the linear Hamiltonian of usual photons.

The explicit form of the DSF in (\ref{DSF}) shows its rational
dependence on the excitation number operator $N$. A very similar DO,
whose DSF $\phi_{\mu}(N)=N/(1+\mu N)$ also has rational type
$N$-dependence, is known as the $\mu$-oscillator of Jannussis~\cite{Jan}.
What is important, DOs with rational nonlinearity principally
differ from the best known one-, two-parameter Fibonacci DOs \cite{AC,BM,Fib}
based on DSFs of exponential type. The latter admit the 3-, 4- and 5-parameter
(exponential type) extensions given in \cite{CCN,MCLD,Bur},
and all the three belong to the Fibonacci class of DOs as well~\cite{GKR}.
On the contrary, DOs with rational nonlinearity of DSF
-- the $\mu$-oscillator and alike -- are not Fibonacci, but
"quasi-Fibonacci" ones~\cite{GKR}.

There are  a few papers exploring diverse aspects of
$\mu$-oscillator and of the related $\mu$-analogs of Bose gas model.
Besides the already mentioned~\cite{Jan,GKR}, let us quote the works
developing $\mu$-Bose gas model in the context of $n$-particle
distributions and correlation function intercepts \cite{GR,GM}, or
from the viewpoint of $\mu$-deformed thermodynamics \cite{RGK}. Most
recently, an application of (the condensate of) $\mu$-Bose gas has
been proposed~\cite{GKKhN} for effective modeling of main properties
of dark matter haloes surrounding dwarf galaxies.

\section{Black-body Radiation at Planck Scale}

In principle, there are different possibilities of describing the
modified radiation spectrum, consisting of the infinite number of
photon modes, corresponding to various frequencies. Most reasonable
point of view, proved in \cite{M09} for the DSR theories of
many-particle systems, suggests to preserve the standard additive
form of a total energy and, therefore, to neglect non-local
interactions among different modes. In practice, we use the
$\kappa$-addition for the identical particles only, i.e. the photons
of the same wave length. Thus, the $\kappa$-addition, mathematically
formulated for the arbitrary distanced particles, induces a
supplemental (non-local) attraction among monochromatic photons,
which reduces their total energy-mass. At the same time, the photon
density turns out here to be of order of Planckian density,
$\kappa^3(\hbar c)^{-3}$.

Under these conditions, we do not consider the soccer-ball problem
of DSR \cite{Hoss}, concerned mainly with the macroscopic bodies,
and discard it due to focusing on the dense (and compact) system,
consisting of the non-interacting subsets of identical light quanta.
However, similarly to the other DSR theories, the terms of
$1/\kappa$-expansion of a total energy-mass would diverge by the
power law with the growing number of photons~\cite{Hoss}, while an
exact expression does not exceed $\kappa$. This is related with a
peculiarity of accounting for supplemental attraction, generated by
the modified energy-momentum addition.

Applying the Planck-scale restriction only to the
subsystem of identical photons with the same cyclic frequency
$\omega$ and the photon (occupation) number $n_\omega$, we write
\begin{equation}\label{Espec1}
E(\{n_\omega\})=\sum\limits_{\omega}E_{\omega,n},\qquad
E_{\omega,n}=\frac{\hbar\omega n_\omega}{1+\hbar\omega(n_\omega-1)/\kappa},
\end{equation}
where summation operation reflects discreteness of the spectrum in a
finite volume $V$; dispersion relation in three space dimensions is
also assumed to be $\omega=c|{\bf k}|$, where ${\bf k}$ is the wave
vector.

We see that $E_{\omega,n}<\hbar\omega n_\omega$ in general. It
physically means a presence of an additional (long-range) attraction
among identical photons, which is growing together with $n_\omega$.
Thus, a photon emission  requires a {\it higher temperature regime},
comparing the temperature with a characteristic energy of the order
of $\kappa$ in magnitude, to overcome the attraction barrier.

On the other hand, the individual energy of photon $\hbar\omega$
should be restricted by condition $\hbar\omega_\mathrm{max}=\kappa$,
because $E_{\omega,n}=\kappa
n_\omega/(n_\omega-1+\kappa/\hbar\omega)>\kappa$ at
$\hbar\omega\gg\kappa$, that is beyond DSR ideology.

We admit that these restrictions can also appear within other
systems of bosons with supplementary attraction, even disconnected
with the (modified)  special relativity theory. Further on, we are
going to describe the thermodynamic properties of such systems.


\textbf{Grand Partition Function.} Formula (\ref{Espec1}) is used
to determine the partition function $\mathcal{Z}^{\rm tot}$ of the photon
gas in volume $V$ within grand canonical ensemble, with temperature
$T=\beta^{-1}$, fugacity $z$, and statistical weights defined by the
Gibbs measure $\exp{(-\beta E_{\omega,n})}$.
Since the photon modes determined by various $\omega$ are viewed as
distinct, the function $\mathcal{Z}^{\rm tot}$ is given as
\begin{equation}\label{Ztot}
\ln\mathcal{Z}^\mathrm{tot}=\sum\limits_{\omega}\ln\mathcal{Z}_\omega,
\end{equation}
\begin{equation}\label{Z1}
\mathcal{Z}_\omega=\rme^{-\beta\kappa}\sum\limits_{n=0}^\infty z^n
\exp{\lb\beta\kappa\frac{1-\mu_\omega}{1+\mu_\omega(n-1)}\rb},
\qquad
\mu_\omega=\frac{\hbar\omega}{\kappa}.
\end{equation}
 Note that the (mode or) $\omega$-dependence of the deformation
parameter $\mu_\omega$ is quite similar to the momentum dependence
of deformation parameter in \cite{Vakar}.

Since the values of $E_{\omega,n}$ at $n_\omega\geq1$ lie within the
energy band of finite width, $\hbar\omega\leq
E_{\omega,n}\leq\kappa$,  the series (\ref{Z1}) diverges at $z=1$.
Thus, the presence of fugacity $z<1$ guarantees the partition
function convergence. Also, $z$ can be determined by negative
chemical potential $\nu=-|\nu|$ so that $z=\exp{(\beta\nu)}$, what
corresponds to a system with effective attraction.

Main peculiarity of calculating (\ref{Z1}) is related with a factor
$\sim1/(1+\mu_\omega n)$ in the Gibbs measure. Its role can be
statistically  evaluated by introducing an order parameter (mean
field):
\begin{equation}\label{sig1}
\sigma(z,x,\mu_\omega)=\frac{\rme^{-x}}{{\cal Z_\omega}}
\sum\limits_{n=0}^\infty \frac{1-\mu_\omega}{1+\mu_\omega(n-1)}
z^n\exp{\lb x\frac{1-\mu_\omega}{1+\mu_\omega(n-1)}\rb} \, .
\end{equation}
Hereafter $x\equiv\beta\kappa$, and $\mathcal{Z}_\omega=
\mathcal{Z}(z,x,\mu_\omega)$ is also assumed.

\begin{figure}
\begin{center}
\includegraphics[width=0.6\linewidth]{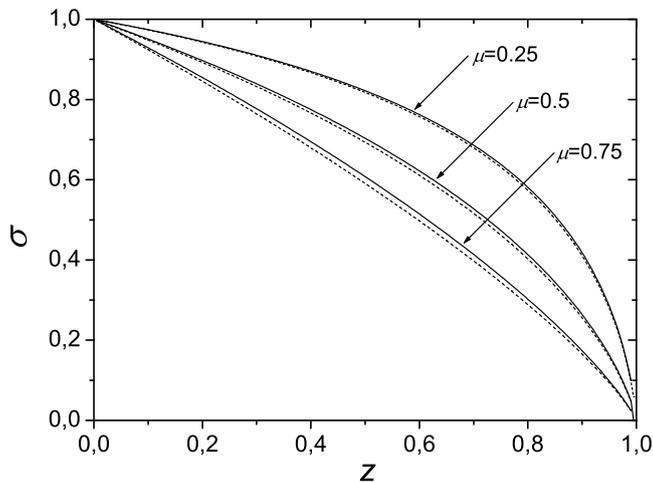}
\end{center}
\vspace*{-7mm}
\caption{Mean field $\sigma$ versus fugacity $z$ at different values
of $\mu$. Solid curves represent the exact dependencies
(\ref{sig1}), computed numerically at $\beta\kappa=0.1$. Dashed
curves, closest to the corresponding solid ones, are obtained
analytically in MFA (\ref{sigeq}) i.e., at $\beta\kappa=0$.}
\end{figure}

One can verify numerically that $\sigma\approx1$ at
$x\gg1$. Non-trivial thermodynamics appears at $x\ll1$ that requires
the temperature $T$ higher than $\kappa$. If $0<\mu<1$, then
$0\leq\sigma\leq1$.
Setting $\mu>1$, there is a region of $(z,x)$ for which $\sigma<0$.

We perform the following calculations assuming that $x<1$ and $0<\mu<1$.
The order parameter existence permits us to evaluate the partition
function and thermodynamic functions in the mean field approximation
(MFA), replacing the factor $(1-\mu_\omega)/(1+\mu_\omega(n-1))$ in
the statistical weights of (\ref{Z1}) by a mean field $\sigma_\omega$. A
self-consistent description requires to formulate simultaneously the
equations for partition function and $\sigma_\omega$:
\begin{eqnarray}
\mathcal{Z}^\mathrm{MFA}_\omega&\equiv&\rme^{-x}\sum\limits_{n=0}^\infty z^n\exp{(x\sigma_\omega)},\\
\sigma_\omega&\equiv&\frac{\rme^{-x}}{\mathcal{Z}^\mathrm{MFA}_\omega}
\sum\limits_{n=0}^\infty \frac{1-\mu_\omega}{1+\mu_\omega(n-1)} z^n \exp{\left(x\sigma_\omega\right)}.
\end{eqnarray}
One obtains
\begin{equation}\label{sigeq}
\mathcal{Z}^\mathrm{MFA}_\omega=\frac{\rme^{x(\sigma_\omega-1)}}{1-z},\qquad
\sigma_\omega=\Lambda\left(z,\mu_\omega\right),
\end{equation}
where the function
\begin{equation}\label{st}
\Lambda(z,\mu)=\frac{(1-\mu)(1-z)}{\mu}\Phi(z,1,\mu^{-1}-1)
\end{equation}
is expressed analytically through the Lerch transcendent
\begin{equation}
\Phi(z,s,a)=\sum\limits_{n=0}^\infty \frac{z^n}{(a+n)^s}
\end{equation}
and visualized by the dashed curves in Fig.~1.
 Note that $\sigma_\omega$ is equal to $\sigma(z,0,\mu_\omega)$,
 and the model equations in MFA are fairly simple to handle.

To relate $\mathcal{Z}^\mathrm{MFA}_\omega$ with $\mathcal{Z}_\omega$ let us expand
$\mathcal{Z}_\omega=\rme^{-x}\sum_{n} z^n
\exp{\left(x\sigma_\omega+xs_{\omega,n}\right)}$
into the series over fluctuations
$s_{\omega,n}=(1-\mu_\omega)/(1+\mu_\omega(n-1))-\sigma_\omega$. A
simple algebra leads to the formula:
\begin{equation}\label{Z3}
\mathcal{Z}_\omega=\mathcal{Z}^\mathrm{MFA}_\omega+
\rme^{x(\sigma_\omega-1)}\sum\limits_{k=1}^\infty
(-x\sigma_\omega)^k \sum\limits_{m=0}^k
\frac{(-\mu_\omega\sigma_\omega)^{-m}(1-\mu_\omega)^m}{m!(k-m)!}
\Phi(z,m,\mu^{-1}_\omega-1),
\end{equation}
where the binomial expansion and the definition of $\Phi(z,s,a)$
have been used. Note that the term indexed by $k=1$ vanishes due to
(\ref{sigeq}).


\textbf{Single-Mode Thermodynamic Functions.}
Computing the main thermodynamic functions, we limit ourselves by
the first non-vanishing terms of expansion in small parameter
$x=\beta\kappa$, depending explicitly on $T$. That yields
\begin{eqnarray}
\mathcal{N}_\omega&=&\left(z\partial_z\ln{\mathcal{Z}^\mathrm{MFA}_\omega}\right)_{\beta,\sigma_\omega}+O(x)\nonumber\\
&=&\frac{z}{1-z}+O(x),\label{N0}\\
U_\omega&=&-\left(\partial_\beta\ln{\mathcal{Z}^\mathrm{MFA}_\omega}\right)_{z,\sigma_\omega}+O(x)\nonumber\\
&=&\kappa[1-\Lambda(z,\mu_\omega)]+O(x),\label{U0}\\
S_\omega&=&\left(\ln{\mathcal{Z}^\mathrm{MFA}_\omega}
-x\partial_x\ln{\mathcal{Z}^\mathrm{MFA}_\omega}\right)_{z,\sigma_\omega}+O(x)
\nonumber\\
&=&-\ln{(1-z)}+O(x).\label{S0}\\
C_\omega&=&\left(x^2\partial^2_x\ln{\mathcal{Z}^{(2)}_\omega}\right)_{z,\sigma_\omega}+O(x^3),
\end{eqnarray}
where \,
$\mathcal{N}_\omega=\left(z\partial_z\ln{\mathcal{Z}_\omega}\right)_{\beta}$,\,
$U_\omega=-\left(\partial_\beta\ln{\mathcal{Z}_\omega}\right)_{z}$
 and $S_\omega=\left(\ln{\mathcal{Z}_\omega}-x\partial_x\ln{\mathcal{Z}_\omega}\right)_{z}$
 are the total mean number of photons, the total internal energy and the entropy,
respectively. These functions are immediately evaluated in MFA.
Numerical tests confirm a good agreement of $\mathcal{N}_\omega(z)$
from (\ref{N0}) with exact dependencies, obtained at $x\ll1$.

The specific heat $C_\omega$ requires the use of a  partition
function in the next to leading-order approximation:
\begin{eqnarray}
\mathcal{Z}^{(2)}_\omega(z,x,\mu_\omega,\sigma_\omega)&=&{\cal
Z}^\mathrm{MFA}_\omega(z,x,\sigma_\omega)
\left[1+F(z,x,\mu_\omega,\sigma_\omega)\right],\\
F(z,x,\mu_\omega,\sigma_\omega)&\equiv&
\frac{x^2}{2}\left[\frac{(1-\mu_\omega)^2(1-z)}{\mu_\omega^2}\Phi(z,2,\mu^{-1}_\omega-1)\right.
\nonumber\\
&&\left.-2\sigma_\omega\Lambda(z,\mu_\omega)+(\sigma_\omega)^2\frac{}{}\right]
+x[\Lambda(z,\mu_\omega)-\sigma_\omega],
\end{eqnarray}
where the function $F$ is assumed to be a small correction.

From $\ln{\mathcal{Z}^{(2)}_\omega}=\ln{\mathcal{Z}^\mathrm{MFA}_\omega}+F$
(linear approximation in $F$), we also find
\begin{equation}\label{C}                  \label{C_omega}
C_\omega=x^2 \upsilon(z,\mu_\omega)+O(x^3),
\end{equation}
where the function $\upsilon(z,\mu)=\partial^2_x F(z,x,\mu,\Lambda(z,\mu))$ is
\begin{equation}\label{var1}
\upsilon(z,\mu)=\frac{(1-\mu)^2(1-z)}{\mu^2}\Phi(z,2,\mu^{-1}-1)-\Lambda^2(z,\mu).
\end{equation}
One can see that it coincides with the variance
\begin{equation}
\left\langle\left(\frac{1-\mu}{1+\mu(n-1)}\right)^2\right\rangle_\mathrm{MFA}
-\left\langle\frac{1-\mu}{1+\mu(n-1)}\right\rangle^2_\mathrm{MFA}.
\end{equation}
Note that $C_\omega\equiv(-\beta^2\partial_\beta
U_\omega)_{z,\sigma_\omega}=0$  results from substituting $U_\omega$
from (\ref{U0}) or taking $x=0$ in~(\ref{C}).

Analyzing, it is useful to compare the dimensionless internal energy
per photon $u_\omega=U_\omega/(\kappa \mathcal{N}_\omega)$, derived
from  (\ref{U0}) and (\ref{N0}), with exact function
\begin{equation}\label{u}
u(z,x,\mu_\omega)=\frac{\mu_\omega\sum_{n=0}^\infty
n[1+\mu_\omega(n-1)]^{-1} w_n(z,x,\mu_\omega)} {\sum_{n=0}^\infty n
w_n(z,x,\mu_\omega)},
\end{equation}
defined by the reduced statistical weights:
\begin{equation}
w_n(z,x,\mu)=z^n\exp{\lb x\frac{1-\mu}{1+\mu(n-1)}\rb}.
\end{equation}

\begin{figure}
\begin{center}
\includegraphics[width=0.49\linewidth]{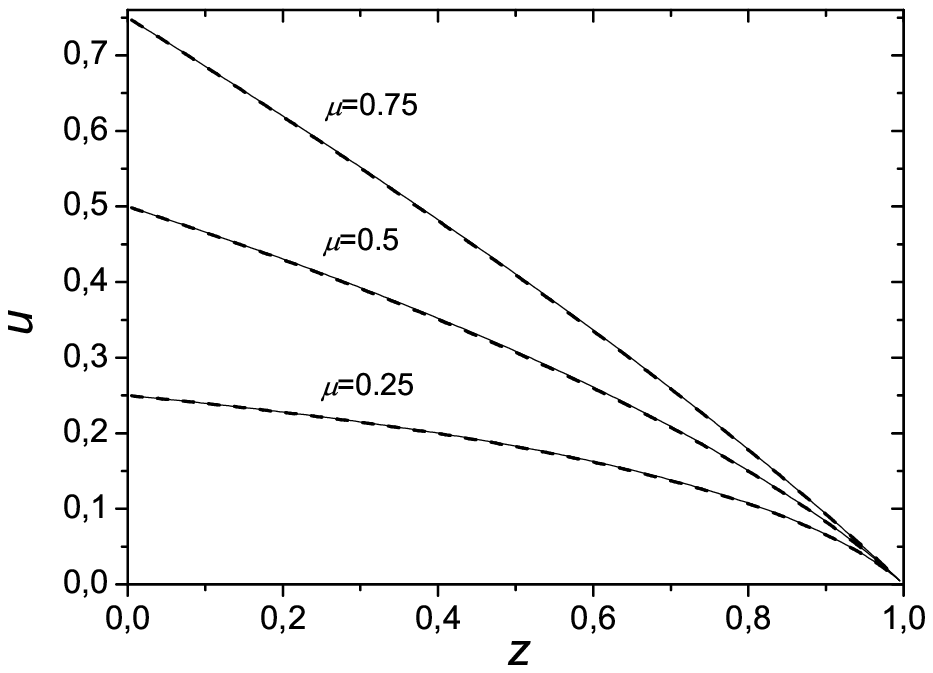}\
\includegraphics[width=0.49\linewidth]{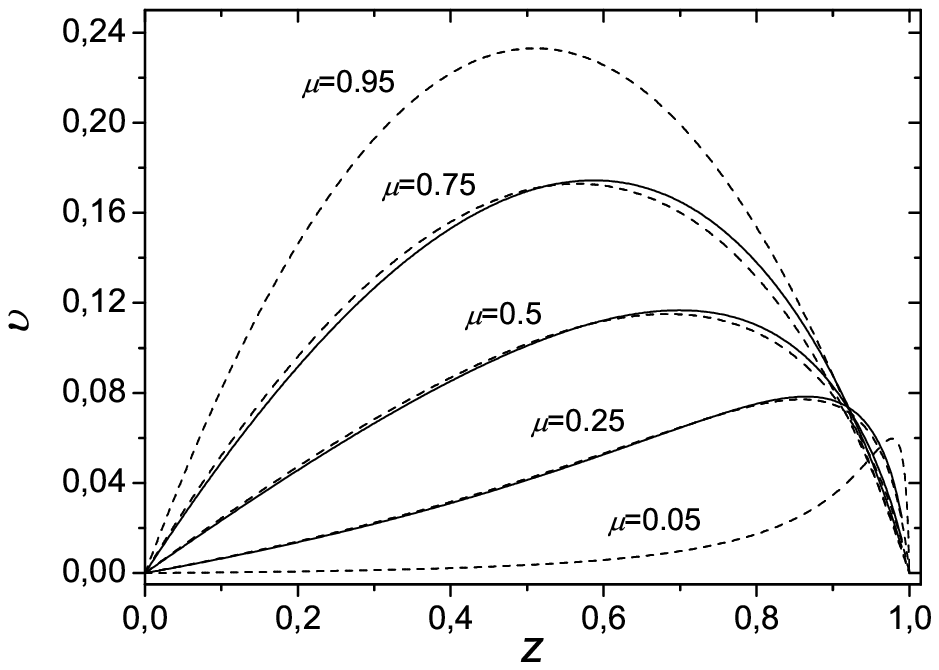}
\end{center}
\vspace*{-10mm}
\caption{\small Dimensionless internal energy $u_\omega$ per photon (left panel)
and variance $\upsilon$ (right panel) vs fugacity $z$ at some $\mu$.
Solid curves represent the exact dependencies (\ref{u}) and (\ref{var1})
at $\beta\kappa=0.1$. Dashed curves are analytically
calculated in MFA.}
\end{figure}

Actually, $u_\omega$ (in MFA) and $u(z,x,\mu_\omega)$ at $x=0.1$ are
seen to coincide in Fig.~2 (left), that confirms validity of our
approach in the regime of high-$T$.  Since the total internal energy
$U_\omega$ increases from 0 to $\kappa$ at $z\to1$, with tending
$\mathcal{N}_\omega\to\infty$, this explains the vanishing $u_\omega$
(and $u(z,x,\mu_\omega)$) at $z\to1$. We also see that the internal
energy per photon $\kappa u_\omega$ tends to $\hbar\omega$ at
$z\to0$.

Now consider the specific heat $C_\omega$ of one-mode subsystem. The
behavior of $\upsilon(z,\mu)$ is seen in Fig.~2 (right), where its
peaks at points $(z_p,\mu_p)$ imply large fluctuations which can be
related to transient processes and are suppressed at $x\ll1$.
Therefore, we expect a presence of two phases in this model.

Note that from (\ref{var1}) it follows:\
$\lim_{\mu\to1}\upsilon(z,\mu)=z(1-z)$.
  This at $\mu\to 1$ yields reflection symmetry with respect
 to the point $z=0.5$.


\textbf{Radiation Characteristics.} Assuming that the volume $V$ is large enough,
we may replace discrete energy spectrum by a continuous one. Then, the number of
quantum states within the interval $(\omega,\omega+\rmd\omega)$ is given by
$(V\omega^2/\pi^2c^3)\,\rmd\omega,$ accounting for the two polarization directions
of photons~\cite{PB11}.

Due to the additivity property of the energies of different modes, the energy of
radiation, accumulated within the frequency interval $(\omega,\omega+\rmd\omega)$ is
$\rmd\mathcal{E}=U_\omega(V\omega^2/\pi^2c^3)\,\rmd\omega.$
Thus, the spectral density of radiation,
\begin{equation}
\frac{1}{4\pi V}\frac{\rmd\mathcal{E}}{\rmd\omega}=
\frac{\kappa^3}{4\pi^3c^3\hbar^2}\,e(z,x,\mu_\omega),
\end{equation}
is determined by the following function derived from $\mathcal{Z}_\omega$:
\begin{equation}\label{e}
e(z,x,\mu)=\mu^3\frac{\sum_{n=0}^\infty n[1+\mu(n-1)]^{-1} w_n(z,x,\mu)}
{\sum_{n=0}^\infty w_n(z,x,\mu)}.
\end{equation}
Since the entities
$\kappa\mu_\omega=\hbar\omega$ and
$x\mu_\omega=\beta\hbar\omega$ are independent of $\kappa$, we have
\begin{equation}
\lim\limits_{\kappa\to\infty}\kappa^3e(z,x,\mu_\omega)=
\frac{(\hbar\omega)^3}{z^{-1}\exp{(\beta\hbar\omega)}-1},
\end{equation}
that leads to the Planck formula for the spectral density of
radiation~\cite{PB11}.

Replacing summation over $\omega$ in (\ref{Ztot}) by integration
with the upper bound $\omega_\mathrm{max}=\kappa/\hbar$ as argued
above, the total energy of radiation is found to be
\begin{equation}\label{Eem}
\mathcal{E}=\frac{V\kappa^4}{\pi^2(\hbar c)^3}\,\varepsilon(z,x).
\end{equation}
Here
%
\begin{equation}\label{e(z,x)}
\varepsilon(z,x)=\int_0^1 e(z,x,\mu)\rmd\mu, \qquad \qquad
\varepsilon_0(z)=\lim\limits_{x\to0}\varepsilon(z,x),
\end{equation}
is the dimensionless emitted energy which is plotted in Fig.~3.

\begin{figure}
\begin{center}
\includegraphics[width=0.6\linewidth]{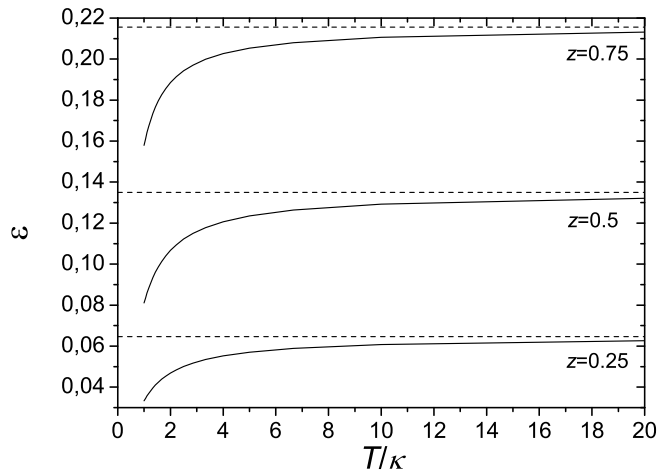}
\end{center}
\vspace*{-10mm}
\caption{\small Temperature dependence of dimensionless emitted
energy $\varepsilon$ from (\ref{e(z,x)}) at various $z$ (numerical
results). Dashed lines correspond to the energy magnitudes at
$T\to\infty$, determined by $\varepsilon_0(z)$ in (\ref{fz}).}
\end{figure}

Analytically, $\varepsilon_0(z)=\int_0^1[1-\Lambda(z,\mu)]\mu^2\rmd\mu$ and equals
\begin{eqnarray}
\varepsilon_0(z)&=&\frac{z}{4}+\frac{z^2}{12}
-z(1-z)\left(\frac{\mathrm{Li}_1(z)}{6}-
\frac{\mathrm{Li}_2(z)}{2}-\mathrm{Li}_3(z)\right)\nonumber\\
&&-z(1-z)\sum\limits_{k=1}^\infty\frac{z^k}{k^4}(k+1)\ln{(k+1)},
\label{fz}
\end{eqnarray}
where $\mathrm{Li}_s(z)$ is the polylogarithm (or Bose)
function: $\mathrm{Li}_s(z)=\sum_{k=1}^\infty z^k/k^s$.

As seen from Fig.~3, at a fixed value $\bar{z}$ of fugacity the
 emitted total energy, taken in units of $V\kappa^4/\pi^2 (\hbar c)^3$,
 is limited by (tends at $T\to\infty$ to) the respective value
 $\varepsilon_0(\bar{z})$.
 The special values of monotonic function
$\varepsilon_0(z)$ are $\varepsilon_0(0)=0$ and
$\varepsilon_0(1)=1/3$, what results from the definition
of $\Lambda(z,\mu)$.

Compare (\ref{Eem}) with the Stefan--Boltzmann law~\cite{PB11}:
$\mathcal{E}_\mathrm{SB}=\pi^2VT^4/(15\hbar^3 c^3)$. While
$\mathcal{E}_\mathrm{SB}\to\infty$ at $T\to\infty$, the DSR based result
predicts a finite value of the radiation energy at arbitrarily large
temperature $T\gg\kappa$.
 As a similar feature, $\mathcal{E}_\mathrm{SB}$ and $\mathcal{E}$ represent the fourth-order
law, defining the dependence of the total emitted energy on $T$ and
$\kappa$ respectively. In the DSR based case, a kind of truncation
of temperature $T$ due to presence of $\kappa$ is observed.

The total number of photons in black-body radiation, the total energy
and the total entropy in MFA are calculated analytically on the base
of (\ref{N0})--(\ref{S0}):
\begin{eqnarray}
\mathcal{N}^\mathrm{MFA}&=&\frac{V\kappa^3}{3\pi^2(\hbar c)^3}\frac{z}{1-z},\label{Ntot}\\
\mathcal{E}^\mathrm{MFA}&=&\frac{V\kappa^4}{\pi^2(\hbar c)^3}\,\varepsilon_0(z),\\
S^\mathrm{MFA}&=&-\frac{V\kappa^3}{3\pi^2(\hbar c)^3}\ln{(1-z)}.
\end{eqnarray}
The equation of state is derived from the relation $\beta
PV=\ln{\mathcal{Z}^\mathrm{tot}}$.
 Using again the expressions in the MFA, one obtains
\begin{equation}\label{eos1}
PV=TS^\mathrm{MFA}-\mathcal{E}^\mathrm{MFA}.
\end{equation}
This formula differs from $P=\mathcal{E}_\mathrm{SB}/(3V)$, which
takes place for the usual ultrarelativistic particles. Unlike that,
expanding the functions (\ref{Ntot})--(\ref{eos1}) into the series
up to the first order in $z$ (when $0<z\ll1$ for
$-\ln{z}\sim\beta\kappa$), we obtain
\begin{eqnarray}
\tilde\mathcal{N}&=&\frac{V\kappa^3}{3\pi^2(\hbar c)^3}\,z,\quad
\tilde\mathcal{E}=\frac{V\kappa^4}{4\pi^2(\hbar c)^3}\,z,\\
\tilde S&=&\frac{V\kappa^3}{3\pi^2(\hbar c)^3}\,z,\quad
\tilde P=\frac{\kappa^4z}{12\pi^2(\hbar c)^3}\left(4\frac{T}{\kappa}-3\right),
\end{eqnarray}
so that the ultrarelativistic particle relations ${\tilde S}\!=\!4{\tilde\mathcal{E}}/(3T)$
 and $\tilde P\!=\!\tilde\mathcal{E}/(3V)$ do hold for $T\!=\!\kappa$.
  Also, ${\tilde S}/{\tilde{\cal N}}\!=\!\mathrm{const}$, in similarity with
the usual photon gas~\cite{PB11}.

\begin{figure}[h]
\begin{minipage}[b]{0.49\linewidth}
\centering
\includegraphics[width=\textwidth]{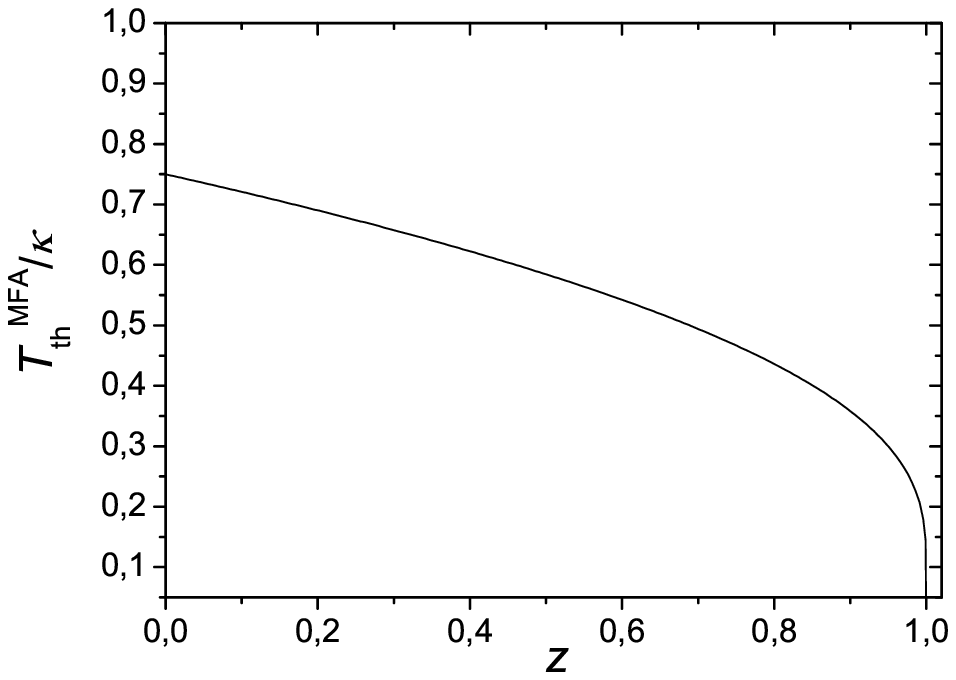}
\vspace*{-12mm}
\caption{\small Temperature threshold of photon emission as function of $z$.}
\label{fig5}
\end{minipage}
\hspace{0.3cm}
\begin{minipage}[b]{0.49\linewidth}
\centering
\includegraphics[width=\textwidth]{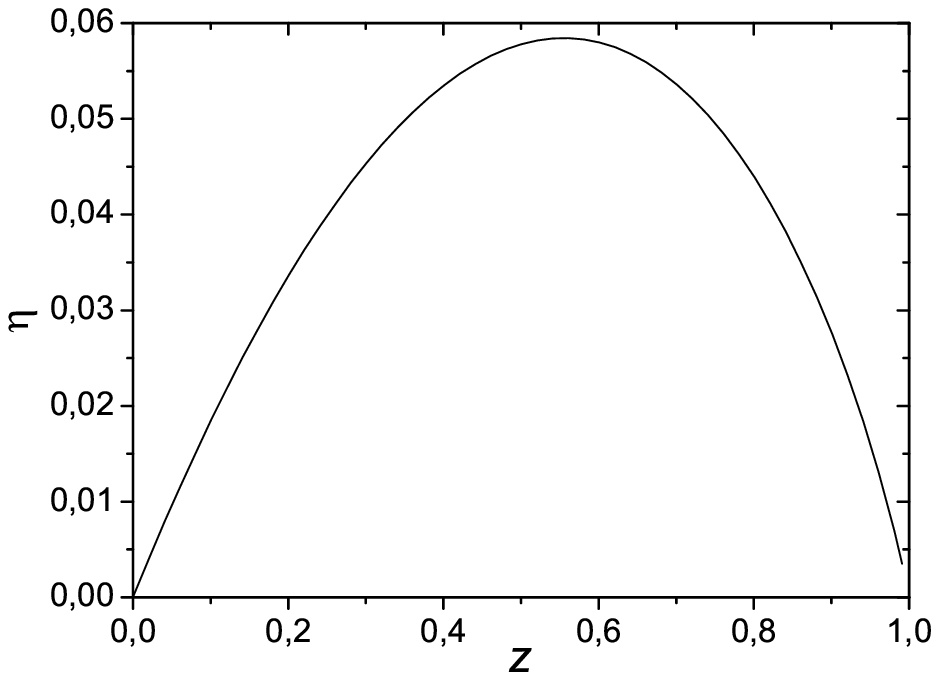}
\vspace*{-12mm}
\caption{\small Function $\eta(z)$  by which the specific heat
$C^\mathrm{MFA}$ depends on $z$.}
\label{fig6}
\end{minipage}
\end{figure}

However, the formulas (\ref{Ntot})--(\ref{eos1}) cannot reproduce
the known expressions for black-body radiation in the limit
$\kappa\to\infty$ because they are obtained in the MFA,
corresponding to the high temperature regime.

Emission ability in the model with effective attraction can be given
by the condition $P\geq0$, that is realized at the temperatures $T$
larger than a threshold $T_\mathrm{th}$. We find the temperature
threshold $T_\mathrm{th}$, when the system starts to radiate, by
requiring $P=0$. Formula (\ref{eos1}) allows us to evaluate
$T^\mathrm{MFA}_\mathrm{th}$ as a function of fugacity $z$.
Analytically, that results in
\begin{equation}
\frac{T^\mathrm{MFA}_\mathrm{th}(z)}{\kappa}=-3\frac{\varepsilon_0(z)}{\ln{(1-z)}},\qquad
\lim\limits_{z\to0} \frac{T^\mathrm{MFA}_\mathrm{th}(z)}{\kappa}=\frac{3}{4},
\end{equation}
what is depicted in Fig.~4.

Integrating, the total specific heat in MFA is obtained from (\ref{C_omega}) as
\begin{equation}
C^\mathrm{MFA}=\frac{V}{T^2}\frac{\kappa^5}{\pi^2(\hbar c)^3}\,\eta(z),
\end{equation}
where we define $\eta(z)=(1-z)\eta_1(z)-(1-z)^2\eta_2(z)$, and
\begin{equation}\label{eta12}
\eta_\alpha(z)=\int_0^1\left(\Phi(z,3-\alpha,\mu^{-1}-1)\right)^\alpha(1-\mu)^2\rmd\mu,\ \
\alpha=1,2.
\end{equation}
Functions $\eta_{1,2}$ can be analytically calculated and are
written out in explicit form in Appendix. The result of calculations
is presented in Fig.~5.

Analyzing, we notice that the specific heat of ordinary black-body
radiation $C\sim VT^3$, while $C^\mathrm{MFA}\sim V\kappa^5T^{-2}$.
However, such rather unusual behavior of $C^\mathrm{MFA}$ is
inherent to the systems with a finite energy band~\cite{PB11}
(similarly to magnetic systems), what has assumed at the beginning.

\section{Conclusion}

In this paper, an original $\kappa$-addition rule inspired by the
DSR has been proposed, exhibiting the crucial role played by the
Planck energy scale $\kappa$ in all our treatment. That rule has
naturally led us to the nonlinear Hamiltonian of self-interacting
one-mode systems of photons. The adopted Hamiltonian, possessing
essentially nonlinear (rational) dependence on the excitation number
operator and combined with the assumption of independence of
different modes, was taken as a starting point for the evaluation,
within the framework of DSR, of main thermodynamic quantities of
black-body radiation. Clearly, the presence of the scale $\kappa$
manifested its importance in our main results on the thermodynamic
characteristics and their physical implications.

First of all, the energy of one-mode subsystem has the property that
it lies entirely in the band of finite width, and the upper bound is
determined by $\kappa$. This property influences all the other
thermodynamic functions.

Next, as follows from Eq.~(48) and is clearly shown in Fig.~4,
within our approach a kind of {\it threshold temperature} $T_{\rm
th}$ (depending on $z$) appears: it implies that just above these
values $T_{\rm th}(z)$ the radiation is present. The disclosed
property of the DSR-based black-body radiation may have important
consequences and unexpected manifestations.

Also it is worth to emphasize the peculiar behavior of the one-mode
specific heat and the total one (shown respectively in Fig.~2 and
Fig.~5), as well as the unusual dependence on the temperature that
was pointed out in the paragraph above Eq.~(42).

An interesting equation of state is obtained which essentially
differs from what is familiar in the standard physics of black-body
radiation. We hope to explore its implications in a separate work.

It is worth to comment on some works on black-body physics based on
deformed
thermodynamics~\cite{Delgado91,Angelo94,Gupta94,Tsallis95,Ch-Ch,Zhang}
and compare their conclusions with the well-known handbook results
\cite{PB11} and with those presented above.
 In the mentioned papers, main novelty that appears due to deformation,
 consisted in some modification of pre-factors in the inferred versions
 of the Stefan-Boltzmann law. Besides, the Stefan's constant begins to
depend on a parameter of deformation. The Wien displacement law is
still preserved for the deformed Bose gas, though with certain
inclusion of deformation parameter. In \cite{Ch-Ch}, the Planck
formula for the deformed Bose gas is really different from the
ordinary one: there appear some new terms in Planck's formula, which
correspond to the "interactions" among photons. Similar to the case
of ideal Bose gas, the total energy of the deformed Bose gas is
proportional to the fourth power $T^4$. The peculiar feature is that
the Stefan-Boltzmann constant turns out to be effectively reduced by
the deformation.

In general, most of the results presented in our paper differ from
those just mentioned in a principal way, namely what concerns the
energy lying within a finite band, the peculiar behavior of specific
heat, and the existence of threshold temperature for radiation
switching. We hope to develop more specified applications of the
obtained results for description of realistic objects in
astrophysics and for effective modeling in modern cosmology.


\textbf{Acknowledgement.}
This work was partly supported by the National Research Foundation
of Korea Grant funded by the Korean Government
(NRF-2015R1D1A1A01057792) and by Development Fund Foundation,
Gyeongsang National University, 2018. Also, the work was partly
supported by The National Academy of Sciences of Ukraine (project
No. 0117U000237).

\section*{Appendix: Computational Results}

Here we write down functions, defined by (\ref{eta12}), as
\begin{eqnarray}
\eta_1(z)&=&\frac{1}{3}+\frac{z}{30}
+\sum\limits_{n=2}^\infty\left(\frac{n^2+10n+1}{3(n-1)^4}-2\frac{n(n+1)}{(n-1)^5}\ln{n}\right)z^n,
\label{eta1}\\
\eta_2(z)&=&\frac{1}{3}+\frac{z}{6}+\left(4\ln{2}-\frac{79}{30}\right)z^2
+\sum\limits_{n=3}^\infty \left(
\frac{2n}{(n-1)^4}\ln{n}\right.\nonumber\\
&&+2\frac{(n-1)^2}{(n-2)^5}\ln{(n-1)}+\frac{1}{3(n-1)}
-\frac{n+1}{(n-1)^3}\nonumber\\
&&\left.
+\frac{n}{6(n-2)^2}-\frac{n(n-1)}{(n-2)^4}+A_n\right)z^n.
\end{eqnarray}

Coefficient $A_3=0$ and $A_{n\geq4}=\sum_{k=1}^{n-3}I(n-2-k,k)$, where
\begin{eqnarray}
I(n,k)&=&\int_0^1\frac{\mu^2(1-\mu)^2}{(1+\mu n)(1+\mu k)}\rmd\mu\\
&=&\frac{(n+1)^2}{n^4(n-k)}\ln{(n+1)}-\frac{(k+1)^2}{k^4(n-k)}\ln{(k+1)}\nonumber\\
&&+\frac{1}{6n^3k^3}[2k^2n^2+9kn(n+k)+6(n^2+kn+k^2)],\
n\not= k;\\
I(n,n)&=&\frac{n^2+12n+12}{3n^4}-2\frac{(n+1)(n+2)}{n^5}\ln{(n+1)}.
\end{eqnarray}
One finds that $I(n,n)=\lim_{k\to n}I(n,k)$, and $I(n-1,n-1)$ is already used as
the series coefficient in~(\ref{eta1}).


\end{document}